\documentclass[journal]{IEEEtran}

\IEEEoverridecommandlockouts                              

\usepackage{epsfig}
\usepackage{epstopdf}
\usepackage{graphics} 
\usepackage[cmex10]{amsmath} 
\usepackage{amssymb}  
\usepackage{microtype}
\usepackage{bm}
\usepackage{color}

\usepackage{amsthm}
\usepackage{url}

\usepackage{amsmath}
\usepackage{graphicx}
\usepackage{cite}
\usepackage{booktabs}
\usepackage{diagbox}
\usepackage{multirow}

\usepackage{algorithm}
\usepackage{algorithmic}

\theoremstyle{definition}

\newtheorem{definition}{Definition}

\newtheorem{theorem}{Theorem}

\everymath{\displaystyle}

\newcommand{\vect}[1]{\text{vec}\left(#1\right)}
\title{\LARGE \bf Model Order Reduction of Large-Scale Wind Farms: A Data-Driven Approach}

\author{Zilong Gong, Junyu Mao, Adri\`{a} Junyent-Ferr\'{e}, \textit{Senior Member, IEEE}, and Giordano Scarciotti, \textit{Senior Member, IEEE}
\thanks{Zilong Gong, Junyu Mao, Adri\`{a} Junyent-Ferr\'{e} and Giordano Scarciotti are with the Department of Electrical and Electronic Engineering, Imperial College London, London, SW7 2AZ, UK. \{zilong.gong18, junyu.mao18, adria.junyent-ferre, g.scarciotti\}@imperial.ac.uk.}
}%

\begin{document}

\maketitle

\begin{abstract}
This paper proposes a data-driven algorithm for model order reduction (MOR) of large-scale wind farms and studies the effects that the obtained reduced-order model (ROM) has when this is integrated into the power grid. With respect to standard MOR methods, the proposed algorithm has the advantages of having low computational complexity and not requiring any knowledge of the high order model. Using time-domain measurements, the obtained ROM achieves the moment matching conditions at selected interpolation points (frequencies). With respect to the state of the art, the method achieves the so-called \emph{two-sided} moment matching, doubling the accuracy by doubling the interpolated points. The proposed algorithm is validated on a combined model of a 200-turbine wind farm (which is reduced) interconnected to the IEEE 14-bus system (which represents the unreduced study area) by comparing the full-order model and the reduced-order model in terms of their Bode plots, eigenvalues and the point of common coupling voltages in extensive fault scenarios of the integrated power system.
\end{abstract}

\section{Introduction}\label{Introduction}
In recent years, there has been a significant increase in efforts to reduce the dependency on coal, petroleum, natural gas and other fossil fuels in order to overcome the effects of global warming. Society as a whole is motivated to replace traditional synchronous generators with renewable generations. In particular, wind farms, which are considered to be a clean, reliable and constant source of energy, are being deployed worldwide \cite{diaz2020review}. However, it is worth noting that in average more than 200 wind turbine generators (WTGs) are required to replace a single coal-fired power plant \cite{usdata}. As a result, a detailed model of a large-scale wind farm typically contains thousands of differential equations. Such a detailed wind farm model can quickly become too complicated for further analysis, especially when hundreds of scenarios need to be simulated. These simulations are often time-consuming or even infeasible due to limited computational power. A solution to this problem is to find reduced models that capture the main dynamics of the large-scale wind farm in a substantially smaller number of differential equations while preserving the main properties (\emph{e.g.} stability) of the whole interconnected power system. This problem is known as model order reduction (MOR). In the context of power systems, model order reduction is normally applied to a partition of the whole network. The idea is to distinguish between a study area, the description of which is maintained in full detail, and an external area (\emph{e.g.} an off-shore wind farm), consisting of the remaining part of the network, which is linearised and reduced. The study area can be precisely analysed and controlled while still considering the interaction of the interconnection of such area with a less faithful representation of a larger external area. This allows faster simulations of hundreds of scenarios in a fraction of the time, \cite{scarciotti2016low,ali2018model,ali2019trajectory,ali2020model}. 

Several MOR methods have been successfully applied to large-scale wind farms. Historically, the most widely used one is the aggregation method \cite{akhmatov2002aggregate,slootweg2003aggregated,fernandez2008aggregated,fernandez2009equivalent,ali2012wind,zou2015fuzzy,wang2018improved}, which simply assumes that all WTGs operate uniformly at the same wind speed and output power, and replaces all WTGs with a single aggregated WTG. However, given that large-scale wind farms are typically distributed over a large area, it is unrealistic to assume that the WTGs operate under the same conditions. To overcome this, one may use multiple aggregated WTGs to represent WTGs under different operating conditions. However, the implementation of this method in wind farms remains limited. Unlike classical generators, which can be easily classified into coherent groups by means of their physical characteristic (\emph{e.g.} rotor angle), the criteria for classifying the WTGs remains unclear. Furthermore, the operating conditions of the wind farm change continuously, resulting in dynamic coherent groups over time. 

Due to the above limitations, non-coherent based techniques have been considered for the reduction of wind farm models, such as Balanced Truncation (BT) \cite{moore1981principal,gugercin2004survey,antoulas2005approximation}, and Rational Krylov methods (RK) \cite{grimme1997krylov,antoulas2005approximation,druskin2011adaptive}. Some successful applications of BT and RK methods to power systems can be seen, for instance, in \cite{chaniotis2005model, freitas2008gramian,chow2013power,sturk2014coherency}. However, these methods have high computational complexity when reducing detailed wind farm models (\emph{e.g.} some may require solving high dimensional Lyapunov equations). To improve this aspect, faster versions of these algorithms have been proposed, \emph{e.g.} the Alternating Direction Implicit (ADI)-based Balanced Truncation \cite{druskin2011analysis} and the Iterative Rational Krylov Algorithm (IRKA) \cite{gugercin2006rational}. A detailed comparison of these methods is given in \cite{ali2018model}. In addition to the high computational complexity, another weakness of these methods is that they require full information about the model (\emph{i.e.} the system matrices $A$, $B$ and $C$, {see \eqref{system}}), which is not a realistic assumption, especially in the case of large-scale wind farms. Wind farm model parameters are commonly owned by commercial companies and are not made available to the operators and users, who may have access only to a black-box model. Therefore, in practice effective MOR approaches should be \textit{model free} or, in other words, \textit{data-driven}. A successful application of a data-driven approach to power systems is \cite{scarciotti2016low}, in which a benchmark model of 16 classical generators is reduced. In particular, \cite{scarciotti2016low} (see also \cite{scarciotti2017data} for the mathematical development) proposed a \textbf{one-sided} time-domain data-driven MOR algorithm based on moment matching. This approach allows selecting arbitrary ``interpolation points'' in the frequency domain such that the Bode plots of the full-order model (FOM) and of the reduced-order model (ROM) are matched at these particular frequencies. Some critical frequencies, which are important for analysis purposes and are strongly related to the mechanical and electrical characteristics of the power system are the so-called critical modes. Typically, critical modes are indicated by significant peaks in the magnitude of the Bode plot. Therefore, the moment matching approach has the notable advantage of enabling the preservation of the critical modes, thus maintaining some physical meaning while reducing the model. 

In this paper, we propose a fully data-driven time-domain \textbf{two-sided} model reduction method for multiple input, multiple output (MIMO) systems. Moreover, the method is validated on a combined model of a 200-turbine wind farm (which represents the external area) and the IEEE 14-bus system (which represents the study area) in terms of Bode plots, eigenvalues, PCC (Point of Common Coupling) voltages and fault behaviour of the FOM and ROM. With respect to the techniques for model order reduction of wind farms available in the literature, the proposed method has the following key advantages and innovations.
\begin{itemize}
\item \textbf{Data-driven method.} This allows to apply the method even if the parameters of the wind farm are IP-protected and unknown to the user.

\item \textbf{Low computational complexity.} Even when the model is available, the proposed method provides computational advantages with respect to model-based MOR. In this case synthetic data is generated in simulation by accurate or black-box models, and is noise free. We show that the execution of the proposed data-driven method has computational speed which is hundreds of times faster than the equivalent model-based method. 

\item \textbf{Robustness.} In practice, real measured data commonly contains measurement noise. Therefore, effective data-driven algorithms should be robust against noise. In the validation section (Section~\ref{Sec_application}), we show a combination of comparisons with and without noise, to illustrate the advantages of the proposed method in both the real and synthetic data scenarios.

\item \textbf{Two-sided.} The method achieves \textbf{two-sided} moment matching. By ``two-sided'' we mean that the number of interpolation points is \textit{doubled} with respect to previously available methods (e.g.\cite{scarciotti2016low}). The implication is that for the same order, the ROM that we produce is as accurate as one with double its order. Alternatively, we can construct a ROM with the same number of interpolation points but with \textit{half} the order. 
\end{itemize} 

The rest of the paper is organised as follows. In Section~\ref{Sec_preliminary}, some preliminaries on moment matching are given. With a specific focus on MIMO systems, we introduce the moment matching conditions along tangential directions. In Section~\ref{Sec_DDMOR}, we propose data-driven algorithms for the MOR of MIMO systems. In Section~\ref{Sec_WFModel}, we describe the combined model of the wind farm and the IEEE 14-bus system, the nonlinear model of the study area and the linearised model of the external area. In Section~\ref{Sec_DesignROM} the external area is reduced and compared with the FOM in terms of the Bode plot, while in Section~\ref{Sec_Robust} the robustness of the data-driven algorithms is assessed under noise conditions. In Section~\ref{Sec_validation} the ROM is reconnected to the study area and validated in terms of eigenvalues and the PCC voltage. In Section~\ref{Sec_time}, the performance of the data-driven MOR method is evaluated by comparing the time required to construct the ROM and the time to execute a simulation. In Section~\ref{Sec_fault}, we analyse the fault behaviour of the FOM and ROM. In section~\ref{Sec_conclusion}, we provide some concluding remarks. Finally, an {Appendix} provides the proofs of the theorems and an external reference containing the full detailed model, additional figures, and a working demo.

\emph{Notation}: The symbols $\mathbb{R}$ and $\mathbb{C}$ denote the sets of real and complex numbers while the symbol $\mathbb{Z}$ denotes the set of integer numbers. The identity matrix is denoted by $I$. The transpose of the matrix $A$ is denoted by $A^{\top}$ while the vectorization of the matrix $A$ is denoted by $\vect A$. $\sigma(A)$ denotes the spectrum of the matrix $A\in \mathbb{R}^{n\times n}$. $\lVert \cdot \lVert$ denotes the induced Euclidean  matrix norm and $\delta_0(t)$ denotes the delta-Dirac impulse. $0_{m\times n}$ represents the $m\times n$ zero matrix. The symbol $e_j$ denotes the $j$-th vector of the standard basis (\emph{i.e. $[\underbrace{0 \cdots 0, 1}_j, 0 \cdots 0]^{\top}$}). 

\section{Preliminaries - Model-based Moment Matching}\label{Sec_preliminary}

In this section we recall the model-based moment matching method for linear MIMO systems. We first recall a family of ROMs matching $\nu$ so-called right interpolating conditions, then a family of ROMs matching $\nu$ so-called left interpolating conditions, and then show how to obtain a ROM that matches $2\nu$ interpolating conditions.

\subsection{Right Interpolation $\equiv$ Direct Interconnection}
Consider a linear MIMO continuous-time system
\begin{equation}
\dot{x}=Ax+Bu, \ \ \ \ \ y=Cx,\label{system}
\end{equation}
where $x(t)\in \mathbb{R}^{n}$ is the state, $u(t)\in \mathbb{R}^{m}$ is the input, $y(t)\in \mathbb{R}^{p}$ is the output, $A\in \mathbb{R}^{n\times n}$, $B\in \mathbb{R}^{n\times m}$, $C\in \mathbb{R}^{p\times n}$ are the system matrices. Without loss of generality system~\eqref{system} is assumed to be the minimal realisation of the transfer matrix $W(s) = C(sI-A)^{-1}B$. We now introduce the definition of moments.

\begin{definition}
\label{D1}
Let $s_i\in \mathbb{C}$\textbackslash$\sigma(A)$ and $l_i\in \mathbb{R}^{m}$. The \textit{$k$-moment of system~\eqref{system} along $l_i$} is a vector defined as $\eta_k(s_i,l_i)=\frac{(-1)^k}{k!}\left[\frac{d^k}{ds_i^k}W(s_i)\right]l_i$, for $k\ge 0$ integer.
\end{definition}

The quantities $s_i$ are called interpolation points, while the quantities $l_i$ are called right tangential directions. 

The idea of MOR by moment matching is to construct a ROM with transfer function $\hat{W}(s)$ (and $k$-moment along $l_i$ called $\hat{\eta}_k(s_i,l_i)$) that, for $\nu$ interpolation points $s_i$, satisfies the right side moment matching condition, namely
\begin{equation}
\eta_k(s_i,l_i)=\hat{\eta}_k(s_i,l_i), \qquad \text{for } i=1,...,\nu.\label{condition1}
\end{equation}

\begin{figure}[thpb]
\centering
\includegraphics[width=0.8\columnwidth]{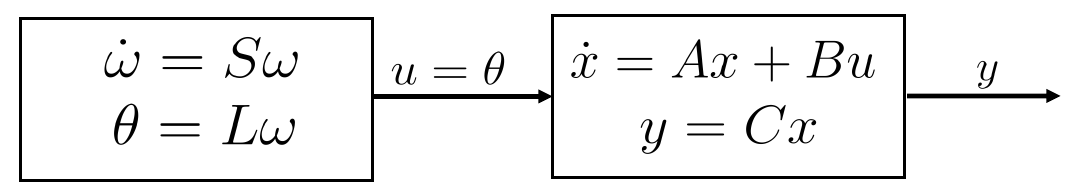}
\caption{The diagram of the direct interconnection.}
\label{fig-dir}
\end{figure}

This problem is readily solved as follows: let $S\in\mathbb{R}^{\nu\times \nu}$ be any matrix with characteristic and minimal polynomials $p(s)=\prod_{i=1}^{\bar{k}}(s-s_i)^{k_i}$, where $\nu=\sum_{i=1}^{\bar{k}}k_i$, and $L=[l_1, l_2, ..., l_\nu]\in \mathbb{R}^{m\times \nu}$, any matrix such that $(S,L)$ is an observable pair. Then, the family of models 
\begin{equation}
  \dot{\xi}=(S-GL)\xi+Gu, \ \ \ \ \psi=C\Pi\xi,  \label{rom}
\end{equation}
achieves moment matching at the interpolation points encoded by $S$ along the directions encoded by $L$, for any $G\in\mathbb{R}^{\nu\times m}$ such that $\sigma(S)\cap\sigma(S-GL)=\emptyset$, as long as $L$ is selected as in \cite[Section 2.4]{shakib2023time} (\textit{e.g.} $l_i=l\ne0_{m\times 1}$ for all $i$). The quantity $\Pi$ in (\ref{rom}) is the unique solution of the Sylvester equation
\begin{equation}
         A\Pi+BL=\Pi S. \label{syl}
\end{equation} 
It is important to notice for the sake of the data-driven algorithms that we develop in this paper, that the quantity $C\Pi \omega(t)$ is the steady-state output $y_{ss}$ of system~(\ref{system}) interconnected (see Fig.~\ref{fig-dir}) with the input produced by the generator
\begin{equation}
 \dot{\omega}=S\omega,  \ \ \ \ \ u=L\omega. \label{signal}
\end{equation}

\subsection{Left Interpolation $\equiv$ Swapped Interconnection}

The same approach can be developed for the so-called left-side interpolation data. 

\begin{definition}
\label{D2}
Let $q_i\in \mathbb{C}$\textbackslash$\sigma(A)$ (interpolation points) and $r_i\in \mathbb{R}^{1\times m}$ (left tangential directions). The \textit{$k$-moment of system \eqref{system} along $r_i$} is a vector defined as $\mu_k(q_i,r_i)=r_i\frac{(-1)^k}{k!}\left[\frac{d^k}{dq_i^k}W(q_i)\right]$, for $k\ge 0$ integer.
\end{definition}

Similarly to before, we now want to find a ROM with moments $\hat{\mu}_k(q_i,r_i)$, such that
\begin{equation}
\mu_k(q_i,r_i)=\hat{\mu}_k(q_i,r_i), \qquad \text{for } i=1,...,\nu.\label{condition2}
\end{equation}

Also this problem is readily solved: let $Q\in\mathbb{R}^{\nu\times \nu}$ be any matrix with characteristic and minimal polynomials $p(q)=\prod_{i=1}^{\bar{k}}(q-q_i)^{k_i}$, where $\nu=\sum_{i=1}^{\bar{k}}k_i$, and $R=[r_1^\top, r_2^\top, ..., r_\nu^\top]^\top\in \mathbb{R}^{\nu\times p}$, any matrix such that $(Q,R)$ is a reachable pair. Then, the family of models 
\begin{equation}
  \dot{\xi}=(Q-RH)\xi+\Upsilon Bu, \ \ \ \ \psi=H\xi,  \label{rom2}
\end{equation}
achieve moment matching at the interpolation points encoded by $Q$ along the direction encoded by $R$, for any $H\in\mathbb{R}^{p\times \nu}$ such that $\sigma(Q)\cap\sigma(Q-RH)=\emptyset$, as long as $R^\top$ is selected as in \cite[Section 2.4]{shakib2023time} (\textit{e.g.} $r_i=r\ne0_{1\times p}$ for all $i$). The quantity $\Upsilon$ in (\ref{rom2}) is the unique solution of the Sylvester equation
\begin{equation}
 \Upsilon A+RC=Q\Upsilon, \label{syl2}
\end{equation}

\begin{figure}[thpb]
\centering
\includegraphics[width=\columnwidth]{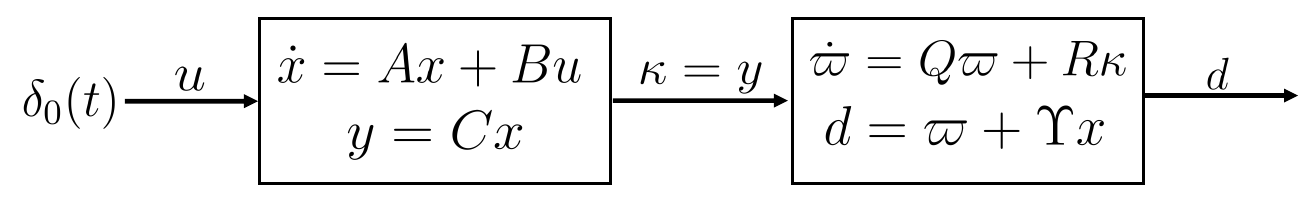}
\caption{The diagram of the swapped interconnection.}
\label{fig-swa}
\end{figure}

It is important to notice (see \cite{mao2022data}) for the sake of the data-driven algorithms that we develop in this paper, that \textbf{for SISO systems} the quantity $e^{Qt}\Upsilon B$ is the output $d$ or steady-state of $\varpi$ of the system
\begin{equation}
 \dot{\varpi}=Q\varpi+Ry,  \ \ \ \ \ d=\varpi+\Upsilon x, \label{signal2}
\end{equation}
that filters the output impulse response of system~(\ref{system}), that is $d(t) = \varpi_{ss}(t) = e^{Qt}\Upsilon B$ (see Fig.~\ref{fig-swa}). In this paper, we prove these properties \textbf{for MIMO systems}.

\subsection{Concurrent Left and Right Interpolation $\equiv$ Two-sided Interconnection}\label{Sec_twosided}

\begin{figure}[thpb]
\centering
\includegraphics[width=\columnwidth]{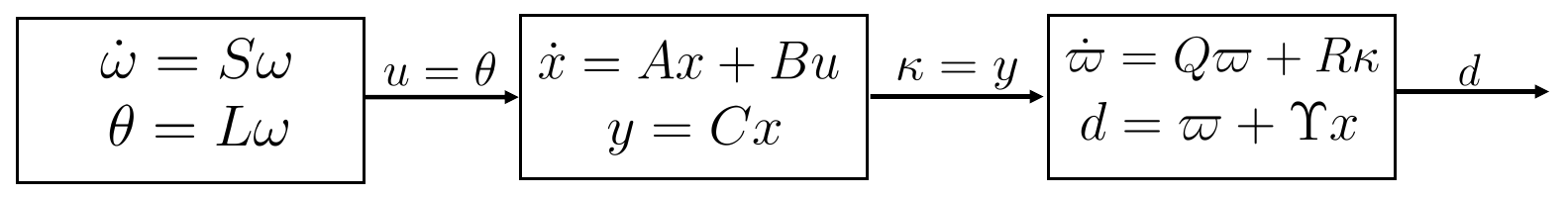}
\caption{The diagram of the two-sided interconnection.}
\label{fig-two}
\end{figure}

The ROM~(\ref{rom}) [(\ref{rom2}) respectively] are of order $\nu$ and match $\nu$ right [left] interpolation condition while they have a free $G$ [$H$]. It is possible to select $G$ [$H$] in such a way that also the left [right] interpolation conditions are satisfied (assuming that these conditions are different, \textit{i.e.} $\sigma(S)\cap\sigma(Q)=\emptyset$). In particular, the ROM~(\ref{rom}) [(\ref{rom2})] with
$$
G=(\Upsilon\Pi)^{-1}\Upsilon B, \qquad [H=C\Pi(\Upsilon\Pi)^{-1}]
$$ 
matches both left and right interpolation data (see \cite{ionescu2015two}). Thus the obtained reduced-order model of order $\nu$ matches $2\nu$ moments. It is important to notice for the sake of the data-driven algorithms that this setup corresponds to the two-sided interconnection shown in Fig.~\ref{fig-two} for which the steady-state property $d_{ss}(t)-\varpi_{ss}(t) = \Upsilon\Pi \omega(t)$ holds.

The idea of the method presented in this paper is to use time series of the outputs of these interconnections to generate a two-sided reduced-order model without solving \eqref{syl} and \eqref{syl2} or using  $A$, $B$ and $C$.

\section{Data-Driven Model Order Reduction}\label{Sec_DDMOR}

One can see that the ROMs proposed in \eqref{rom} and \eqref{rom2} rely on the unique solutions of the Sylvester equations \eqref{syl} and \eqref{syl2} respectively. Since these Sylvester equations involve $n \times \nu$ conditions, if $n$ is large they are computationally challenging to solve. This results in high computational complexity when implementing this approach to wind farms which are commonly large-scale. Furthermore, as mentioned in Section~\ref{Introduction}, detailed information of the system matrices (\emph{i.e.} $A$, $B$ and $C$) of large-scale wind farms may not be available. Thus, we are motivated to find a data-driven MOR approach that requires input-output data only. 

A data-driven algorithm for two-sided MOR of SISO systems has been proposed in \cite{mao2024data}. However, large-scale wind farms are commonly modeled with multiple inputs and multiple outputs (\emph{i.e.} voltages and currents in \emph{dq} framework). Thus, a MIMO method is required. The main difficulty of the MIMO case is that we cannot directly match all quantities of the transfer matrix $W(s)$ because for a matrix $W(s)$ of dimension $m$ (inputs) by $p$ (outputs) and $\nu$ interpolation points, then $mp\nu$ interpolation conditions need to be satisfied, whereas the degrees of freedom of $C\Pi$ are only $p\nu$. Therefore, we merge the $mp\nu$ conditions into just $p\nu$ by using tangential directions as in Definition~\ref{D1} and Definition~\ref{D2}. A data-driven algorithm for \textit{one-sided} MOR of MIMO systems was presented in \cite{scarciotti2016low}. Here we propose a novel algorithm for \textit{two-sided} MOR of MIMO systems, thus producing ROMs with double the accuracy of the method\footnote{There is also a difference in the MIMO aspect with respect to \cite{scarciotti2016low}. In \cite{scarciotti2016low}, a heuristic was used to determine $L$, producing only \textit{approximate} one-sided moment matching. Here, $L$ and $R$ are selected based on the recent results in \cite{shakib2023time}. The resulting ROM is not an approximation.} given in \cite{scarciotti2016low}.

From Section~\ref{Sec_preliminary} we have seen that a reduced-order model by moment matching that achieved two-sided interpolation can be constructed as 
\begin{equation}
  \dot{\xi}=(S-(\Upsilon\Pi)^{-1}\Upsilon BL)\xi+(\Upsilon\Pi)^{-1}\Upsilon Bu, \ \ \ \ \psi=C\Pi\xi,  \label{rom_twos}
\end{equation}
if we have the matrices $C\Pi$, $\Upsilon B$ and $\Upsilon\Pi$. Therefore, the data-driven approach consists in finding accurate approximations $\widetilde{C\Pi}$, $\widetilde{\Upsilon B}$ and $\widetilde{\Upsilon\Pi}$ from data.

\subsection{Constructing $\widetilde{C\Pi}$}

Consider the experimental setup in Fig.~\ref{fig-dir} or Fig.~\ref{fig-two} in which the system is excited with inputs produced by the generator \eqref{signal}. Samples of the output $y(t_i)$ of the system are collected together with $\omega(t_i)$ (the latter can be collected or computed exactly). Then the following theorem (see Appendix A for the proof) provides a formula to estimate $C\Pi$. The theorem provides convergence guarantees for Algorithm~\ref{algCPai}.

\begin{theorem}
Consider the interconnection of system \eqref{system} and the signal generator \eqref{signal}. Assume $\sigma(A)\subset\mathbb{C}_{<0}$, $\sigma(S)\subset\mathbb{C}_{0}$, $S$ has simple eigenvalues and the triple $(L, S, \omega(0))$ is minimal.
For $\widetilde{\nu}\ge \nu$, define the time-snapshots $\widetilde{R}_k\in \mathbb{R}^{\widetilde{\nu}\times \nu}$ and $\widetilde{E}_k^j\in \mathbb{R}^{\widetilde{\nu}}$ as
$$
\begin{aligned}
\widetilde{R}_k&=[\omega(t_{k-\widetilde{\nu}+1}) \cdots \omega(t_{k-1}) \ \omega(t_{k})]^{\top},\\
\widetilde{E}_k^j&=[y^j(t_{k-\widetilde{\nu}}+1) \cdots y^j(t_{k-1}) \ y^j(t_{k})]^{\top},
\end{aligned}
$$
where $y^j(t_{i})$ is the $j$-th row of $y(t_i)$. Denote the $j$-th row of $C$ as $C^j$. Then, 
\begin{equation}
\vect{\widetilde{C^j\Pi}_k}=(\widetilde{R}_k^{\top}\widetilde{R}_k)^{-1}\widetilde{R}_k^{\top}\widetilde{E}_k^j.\label{Cpai}
\end{equation}
is an \textit{online estimate} of $C^j \Pi$, which is the $j$-th row of $C\Pi$, namely there exists a sequence $\{t_k\}$ such that $\widetilde{R}_k^\top \widetilde{R}_k$ is full rank and $\lim\limits_{k\to \infty}\widetilde{C^j\Pi}_k=C^j\Pi$.
\label{direct_dd}
\end{theorem}

\begin{algorithm}
	\renewcommand{\algorithmicrequire}{\textbf{Input:}}
	\caption{Approximate $C\Pi$}
	\label{algCPai}
	\begin{algorithmic}[1]
        \REQUIRE A sufficiently large integer $k_0\in \mathbb{Z}_{>0}$ and sufficiently small error thresholds $\eta_{C\Pi}^j>0$, with $j=1,...,p$. Let $\widetilde{\nu}=\nu$ (increase $\widetilde{\nu}$ sufficiently for robustness).
        \FOR{$j=1$ to $p$}
        \STATE Let $k=k_0$.
		\STATE Construct the matrices $\widetilde{R}_k$ and $\widetilde{E}_k^j$.
		\IF{rank$(\widetilde{R}_k)=\nu$} 
        \STATE Compute $\widetilde{C^j\Pi}_k$ from equation \eqref{Cpai}.
          \IF{$\frac{||\widetilde{C^j\Pi}_k-\widetilde{C^j\Pi}_{k-1}||}{t_k-t_{k-1}}>{\eta_{C\Pi}^j}$}
          \STATE $k=k+1$ go to 3.
          \ENDIF
        \ELSE 
        \STATE Increase $\widetilde{\nu}$ go to 3.
        \ENDIF
        \RETURN $\widetilde{C^j\Pi}_k$.
        \ENDFOR
        \RETURN $\widetilde{C\Pi}_k$.
	\end{algorithmic}  
\end{algorithm}

Algorithm~\ref{algCPai} can be made robust against measurement noise. It is enough to pick a larger $\widetilde{\nu}$. This variant gives the optimal estimation (i.e. maximum
likelihood estimation) of $C^j\Pi$ when the measurements
are affected by Gaussian additive noise.

\subsection{Constructing $\widetilde{\Upsilon B}$}

Consider the experimental setup in Fig.~\ref{fig-swa} in which the output impulse response of system~(\ref{system}) is filtered by (\ref{signal2}). Samples of the filtered state $\varpi(t_i)$ are collected together with the quantities $e^{Qt_i}$ (these can be computed exactly). Then the following theorem (see Appendix A for the proof) provides a formula to estimate $\Upsilon B$. 

\begin{theorem}
Consider the interconnection of system~\eqref{system} and ~\eqref{signal2}. Assume $\sigma(A)\subset\mathbb{C}_{<0}$, $\sigma(Q)\subset\mathbb{C}_{0}$, $Q$ has simple eigenvalues and the pair $(Q, R)$ is reachable. Denote the $j$-th column of $B$ as $B^j$. Let the initial conditions $x(0)=0$ and $\varpi(0)=0$, and the input of the $j$-th experiment $u_j=e_j\delta_0(t)$. Then
\begin{equation}
\widetilde{\Upsilon B}^j_k:=e^{-Q t_k} \varpi(t_k)
\label{YB}
\end{equation}
is an online estimate of $\Upsilon B^j$ with the following asymptotic property: there exists a time sequence $\left\{ t_k \right\}$ such that $\lim\limits_{k\to \infty}\widetilde{\Upsilon B}^j_k=\Upsilon B^j$.
\label{swapped_dd}
\end{theorem}

Equation~\eqref{YB} can be made robust against measurement noise. To this end we replace~\eqref{YB} with 
\begin{equation}
\vect{\widetilde{\Upsilon B}^j_k}=(\widetilde{P}_k^{\top}\widetilde{P}_k)^{-1}\widetilde{P}_k^{\top}\widetilde{O}_k^j,\label{YBls}
\end{equation}
where $\widetilde{O}_k^j\in \mathbb{R}^{\widetilde{q}\nu}$ and $\widetilde{P}_k\in \mathbb{R}^{\widetilde{q}\nu\times \nu}$ are defined as
$$
\begin{aligned}
\widetilde{O}_k^j&=[\varpi(t_{k-\widetilde{q}+1})^{\top} \cdots \varpi(t_{k-1})^{\top} \ \varpi(t_{k})^{\top}]^{\top},\\
\widetilde{P}_k&=[e^{Q^{\top}t_{k-\widetilde{q}+1}} \cdots e^{Q^{\top}t_{k-1}} \ e^{Q^{\top}t_{k}}]^{{\top}},
\end{aligned}
$$
and note that $\widetilde{P}_k$ is always full column rank. For a sufficiently large number $\widetilde{q}$, this variant gives the optimal estimation (i.e. maximum
likelihood estimation) of $\Upsilon B^j$ when the measurements are affected by Gaussian additive noise. The results of this section are summarised in Algorithm~\ref{algYB}.

\begin{algorithm}
	\renewcommand{\algorithmicrequire}{\textbf{Input:}}
	\caption{Approximate $\Upsilon B$}
	\label{algYB}
	\begin{algorithmic}[1]
        \REQUIRE A sufficiently large integer $k_0\in \mathbb{Z}_{>0}$ and sufficiently small error thresholds $\eta_{\Upsilon B}^j>0$ with $j=1,...,m$. Let $\widetilde{q}=1$ (increase $\widetilde{q}$ sufficiently for robustness).
        \FOR{$j=1$ to $m$}
          \STATE Let $k=k_0$.
          \STATE Produce the data with the input $u_j=e_j\delta_0(t)$.
		 \STATE Construct the matrices $\widetilde{O}_k^j$ and $\widetilde{P}_k$.
           \STATE Compute $\widetilde{\Upsilon B}^j_k$ from equation \eqref{YBls}.
           \IF{$\frac{||\widetilde{\Upsilon B}^j_k-\widetilde{\Upsilon B}^j_{k-1}||}{t_k-t_{k-1}}>{\eta_{\Upsilon B}^j}$}
           \STATE $k=k+1$ go to 4.
           \ENDIF
         \RETURN $\widetilde{\Upsilon B}_k^j$.
       \ENDFOR 
       \RETURN $\widetilde{\Upsilon B}_k$.
	\end{algorithmic}  
\end{algorithm}

\subsection{Constructing $\widetilde{\Upsilon\Pi}$}
Provided with satisfying approximations $\widetilde{C\Pi}_k$ and $\widetilde{\Upsilon B}_k$, we are left with the problem of estimating ${\Upsilon\Pi}$ from data. For SISO systems, \cite{mao2024data} proposed a third algorithm based on an additional experiment and the use of a ``surrogate'' system (see \cite{mao2024data} for details). Here we present an alternative method (for both SISO and MIMO systems) that has the advantage of not requiring a third experiment and algorithm. 

By multiplying equation~(\ref{syl}) from the left by $\Upsilon$, and equation~(\ref{syl2}) from the right by $\Pi$ and matching the resulting common terms $\Upsilon A \Pi$, yield the new equation
$$
Q \Upsilon \Pi - \Upsilon \Pi S = RC \Pi - \Upsilon B L.
$$
Exploiting the approximations $\widetilde{C\Pi}_k$ and $\widetilde{\Upsilon B}_k$, we can find an approximation of $\Upsilon\Pi$ by solving the Sylvester equation 
\begin{equation}
\label{eq-ypiSyl}
Q \widetilde{\Upsilon\Pi}_k - \widetilde{\Upsilon\Pi}_k S = R \widetilde{C\Pi}_k - \widetilde{\Upsilon B}_k L
\end{equation}
with respect to $\widetilde{\Upsilon\Pi}_k$. This equation has a unique solution as long as $Q$ and $S$ have no common eigenvalues. Since $Q$ and $S$ are arbitrary, this is not a restriction. Moreover, note that differently from the Sylvester equations \eqref{syl} and \eqref{syl2} that correspond to $n$ by $\nu$ and $\nu$ by $n$ scalar equations, respectively, the Sylvester equation \eqref{eq-ypiSyl} only involves $\nu$ by $\nu$ scalar equations. Since $\nu \ll n$, the computational complexity of solving \eqref{eq-ypiSyl} is small and does not need to be avoided as it was the case for \eqref{syl} and \eqref{syl2}.

\subsection{Algorithmic Overview}
\begin{figure}[thpb]
      \centering
      \includegraphics[width=0.75\columnwidth]{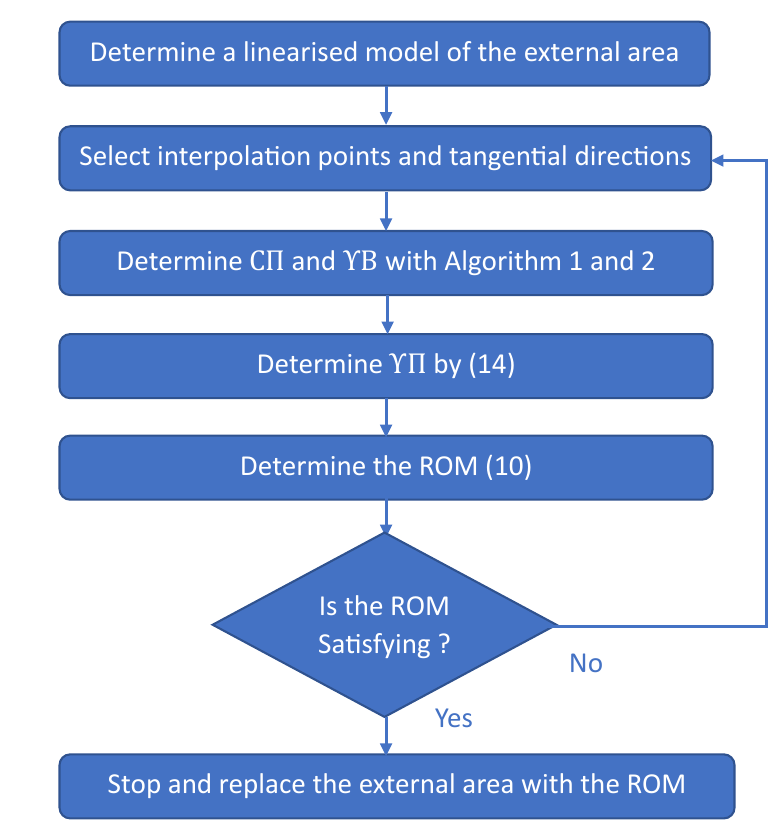}
      \caption{{Algorithmic overview of the method}}
      \label{figure_Alg}
\end{figure}

The complete method is summarised as a flowchart in Fig.~\ref{figure_Alg}. Once the study and external areas are given, we determine a linearised model of the external area. The next step consists in selecting the number and location of the interpolation points and the corresponding tangential directions for the right and left side moment matching. These parameters, which determine the performance of the ROM, can be selected based on user's knowledge of the wind farm or its operating conditions, or by using a fast Fourier transform on available data to determine the frequencies of interest to be interpolated. We generate the data $y(t)$, $\omega(t)$, and $\varpi(t)$, and compute $\widetilde{C\Pi}_k$ and $\widetilde{\Upsilon B}_k$ by Algorithm~\ref{algCPai} and Algorithm~\ref{algYB} respectively. $\widetilde{\Upsilon\Pi}_k$ is obtained by solving (\ref{eq-ypiSyl}). Finally, the two-sided reduced-order model is given by~\eqref{rom_twos}. If the resulting ROM is not satisfactory, we return to the step of selecting the interpolation points and the corresponding tangential directions. Otherwise, we stop. A demo of the method is available in Appendix C.
\section{Application to the wind farms}\label{Sec_application}
In this section we present extensive simulation results to validate the proposed method. After introducing the wind farm model used in this work (Section~\ref{Sec_WFModel}) and the obtained ROM (Section~\ref{Sec_DesignROM}), we first evaluate the robustness of the data-driven algorithms to measurement noise (Section~\ref{Sec_Robust}) and then compare the FOM and the ROM in terms of eigenvalues, PCC voltage, computational time and fault behaviour (Sections~\ref{Sec_validation}, \ref{Sec_time}, \ref{Sec_fault}).

\subsection{Combined system: wind farm and IEEE 14-bus system}\label{Sec_WFModel}
\begin{figure}[thpb]
      \centering
      \includegraphics[width=0.9\columnwidth]{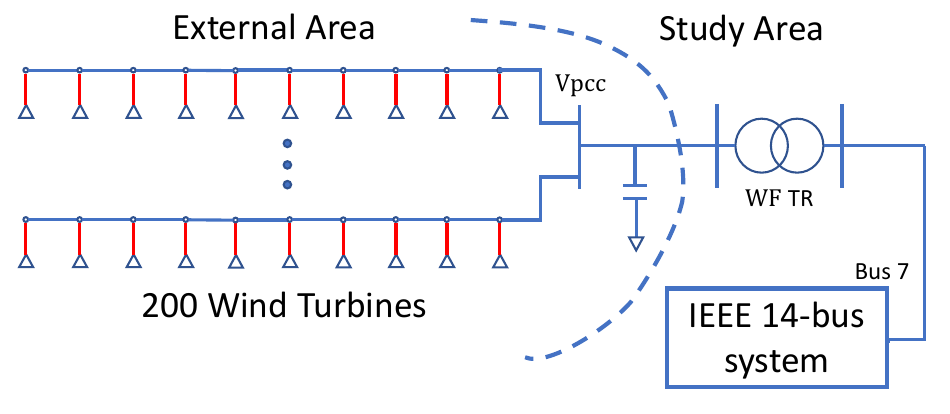}
      \caption{The layout of the combined system. A fully detailed model is provided in \cite{appendix}.}
      \label{figure3}
\end{figure}

To validate our MOR approach, we consider a 200 $\times$ 1MW wind farm of type-4 wind turbine generators (WTGs) in a 50Hz power system. The basic layout of the combined system is shown in Fig.~\ref{figure3}. The wind farm consists of 20 strings, where each string contains 10 WTGs, while the rest of the power system consists of 14 buses, 5 synchronous generators, 11 loads and 20 transmission lines (see \cite{appendix} for more detail). The wind farm is connected to the Bus 7. In Fig.~\ref{figure3}, each triangle represents a WTG connected to a wind turbine transformer (WTT). The dynamic model of a single WTG has order $9$ (see \textit{e.g.} \cite{junyent2011control}), which increases to $17$ once connected with the transformer and cable. The full detailed model, including all the parameters used in this study, is provided in \cite{appendix}. The large-scale wind farm and the IEEE 14-bus system are modelled using Matlab/Simulink. We define the study area and the external area as shown in Fig.~\ref{figure3}. The study area is preserved while the external area is linearised using the linearisation toolbox in {Matlab/Simulink (note that linearisation is common in this context, see \textit{e.g.} \cite{ali2018model,ali2020model}).} The resulting model of the external area has an order of $3402$ with $1592$ conjugate pairs of eigenvalues. This linear FOM of the external area includes two inputs and two outputs, namely the current at the wind farm transformer (WFT) on the \emph{qd}-axis and the PCC voltage on the \emph{qd}-axis, respectively. The FOM of the external area is subsequently reduced by our data-driven approach and reconnected to the study area for comparison and validation against the FOM. In the rest of the section, this combined system is simulated in Simulink using `variable-step' with a max step size of $5.5\times 10^{-6}$. Note that for the purpose of model order reduction, no parameters or system matrices are used. Only the data as defined in the data-driven algorithms are used. The matrices are used only for validation purposes.

\subsection{Design of the reduced-order model}\label{Sec_DesignROM}
Following the development of Section~\ref{Sec_DDMOR}, we implement Algorithm~\ref{algCPai}, Algorithm~\ref{algYB} and solve\footnote{Using the function ``sylvester'' in Matlab.}~\eqref{eq-ypiSyl}, finding estimations of $C\Pi$, $\Upsilon B$ and $\Upsilon\Pi$, respectively. The first step is to determine the interpolation points and the corresponding tangential directions for moment matching. In this example, the interpolation points are selected by spreading them between the frequencies of $0$Hz and $10000$Hz, placing some of these frequencies\footnote{It may not always be possible to arbitrarily select the input of system \eqref{system} as in Fig.~\ref{fig-dir} and Fig.~\ref{fig-swa}, and in reality the input signal may be composed of several unwanted frequencies. For this case, one can use a band-pass filter and apply the algorithms to the filtered data, as suggested for instance in \cite[Remark 2]{scarciotti2017data}.} in $S$ and some in $Q$. Specifically, we used $\pm1\iota$, $\pm20\iota$, $\pm70\iota$, $\pm120\iota$, $\pm300\iota$, $\pm500\iota$, $\pm700\iota$, $\pm900\iota$, $\pm4400\iota$ and $\pm8500\iota$ for Algorithm~\ref{algCPai} and  $\pm10\iota$, $\pm50\iota$, $\pm100\iota$, $\pm200\iota$, $\pm400\iota$, $\pm600\iota$, $\pm800\iota$, $\pm1000\iota$, $\pm7000\iota$ and $\pm10000\iota$ for Algorithm~\ref{algYB}. We select the tangential directions according to \cite{shakib2023time} as $l_i=[1,4]^{\top}$ and $r_i=[4,1]$ for all $i=1,\dots,\nu$ for the direct and swapped moment matching, respectively. The performance of the data-driven algorithms is evaluated by the normalised approximation error defined as
$$
e_{\text{x}}=\frac{||\widetilde{\text{x}}_k-{\text{x}}||} {||\text{x}||},
$$
where $\text{x}$ is any quantity required to construct the ROM (\emph{i.e.} $C\Pi$, $\Upsilon B$ and $\Upsilon\Pi$) and $\widetilde{\text{x}}_k$ is the corresponding approximation given by the data-driven algorithm (\emph{i.e.} $\widetilde{C\Pi}_k$ $\widetilde{\Upsilon B}_k$ and $\widetilde{\Upsilon\Pi}_k$). The system is simulated for 100 seconds to generate sufficient data\footnote{In practice, the required data can be collected by the phase measurement unit (PMU). In this paper, the data is obtained directly from simulations.} for the two algorithms. This yields sufficiently small normalised approximation errors $e_{C\Pi}=2.1\times 10^{-6}$ and $e_{\Upsilon B}=8.8\times 10^{-4}$. 
Once satisfactory approximations of $C\Pi$ and $\Upsilon B$ are obtained, we solve equation~\eqref{eq-ypiSyl} to estimate $\Upsilon\Pi$, yielding a normalised approximation error of $e_{\Upsilon \Pi}=1.2\times 10^{-6}$. Provided with approximations $\widetilde{C\Pi}_k$, $\widetilde{\Upsilon B}_k$ and $\widetilde{\Upsilon\Pi}_k$, the ROM of order $\nu=20$ is given by  \eqref{rom_twos}.

\begin{figure}[htpb]
      \centering
      \includegraphics[width=0.9\columnwidth]{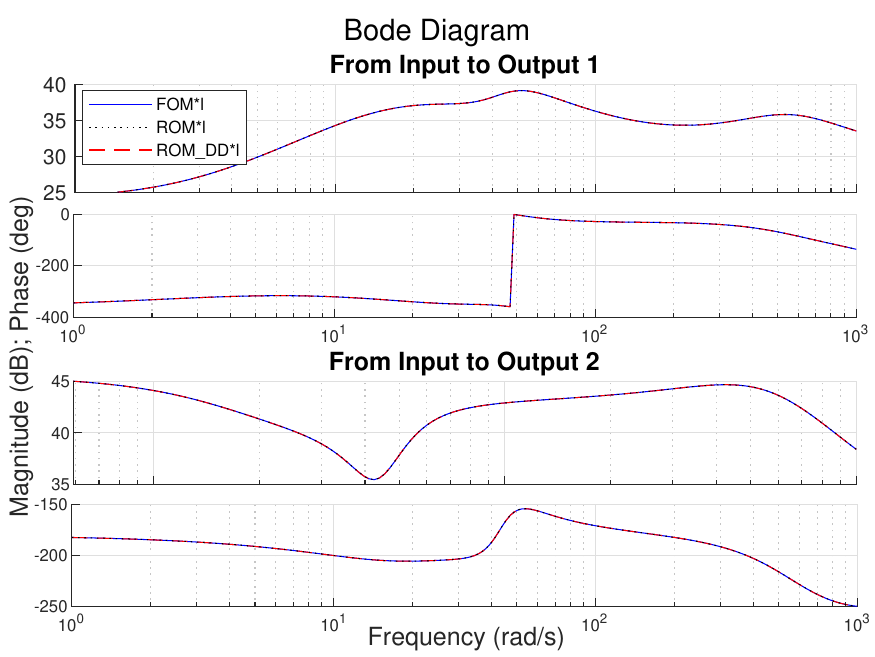}
      \caption{Bode plot of the right side moment matching along the tangential direction $l$. The solid blue line, black dotted line and red dashed line represent the quantities for the FOM, ROM (model-based) and ROM\_DD (data-driven), respectively.}
      \label{sim_direct}
\end{figure}

\begin{figure}[htpb]
      \centering
      \includegraphics[width=0.9\columnwidth]{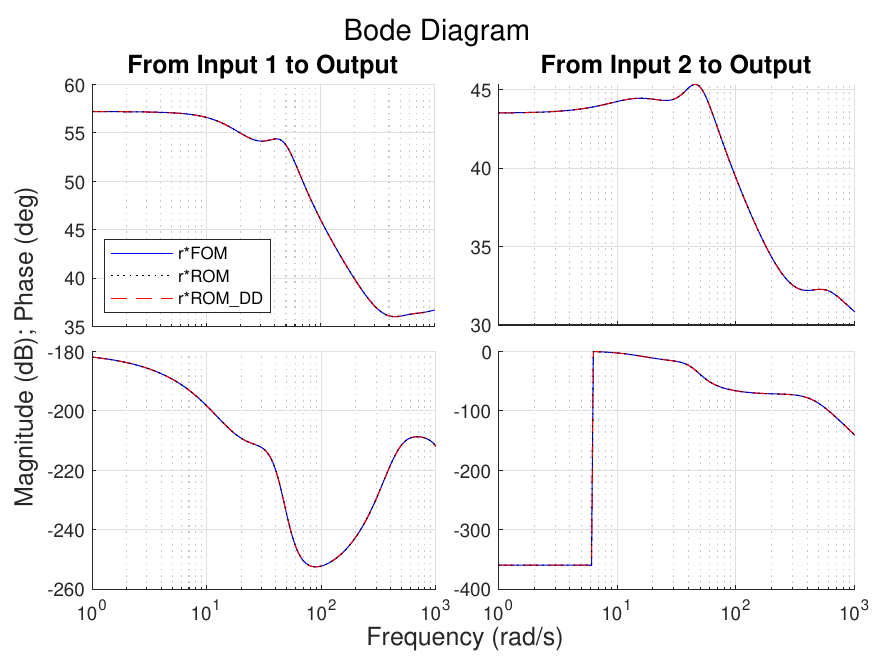}
      \caption{Bode plot of the left side moment matching along the tangential direction $r$. The solid blue line, black dotted line and red dashed line represent the quantities for the FOM, ROM (model-based) and ROM\_DD (data-driven), respectively.}
      \label{sim_swapped}
\end{figure}

\begin{figure}[htpb]
      \centering
      \includegraphics[width=0.9\columnwidth]{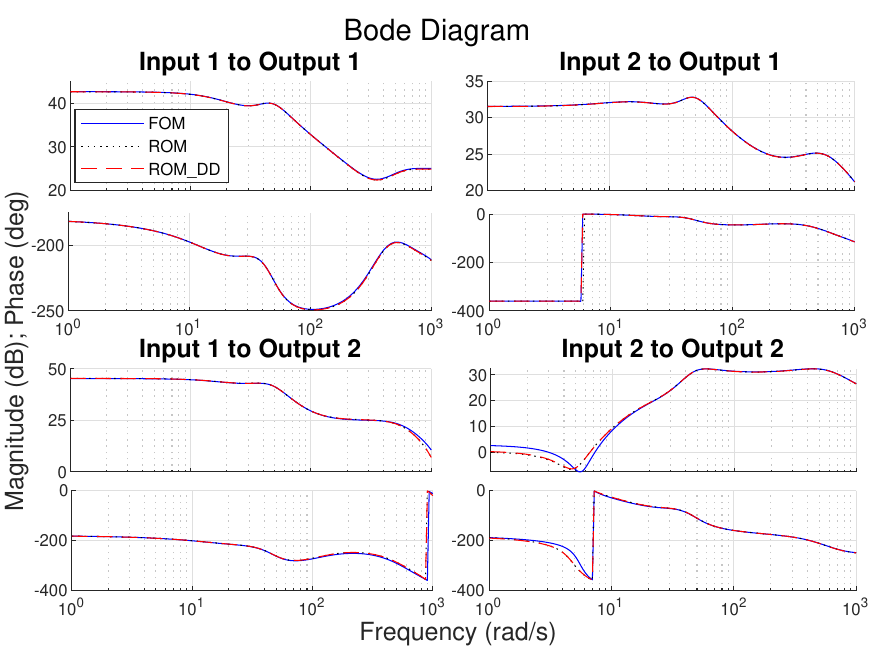}
      \caption{Bode plot of the FOM and ROMs. The solid blue line, black dotted line and red dashed line represent the FOM, ROM (model-based) and ROM\_DD (data-driven), respectively.}
      \label{sim_bode}
\end{figure}

Fig.~\ref{sim_direct} and Fig.~\ref{sim_swapped} compare the Bode plots for the right- and left-side moment matching, respectively. In particular, the blue solid lines represent quantities related to the FOM, the black dotted lines represent quantities related to the model-based ROM and the dashed red lines represent quantities related to the data-driven ROM. The figures show that the selection of the interpolation points is effective and the transfer matrix along the tangential directions is perfectly matched. Hence, the performance of the model-based reduced-order model is fully recovered by the data-driven approach.
In Fig.~\ref{sim_bode}, the Bode plot of the FOM and ROMs (model-based and data-driven) are represented by solid blue, black dotted and red dashed lines, respectively. Note that while there is no guarantee that the transfer matrix is well matched (the guarantee is along the left and right directions only), the figure clearly shows that the transfer matrix itself is well matched across all the considered frequencies. 


\begin{figure}[htpb]
      \centering
      \includegraphics[width=0.9\columnwidth]{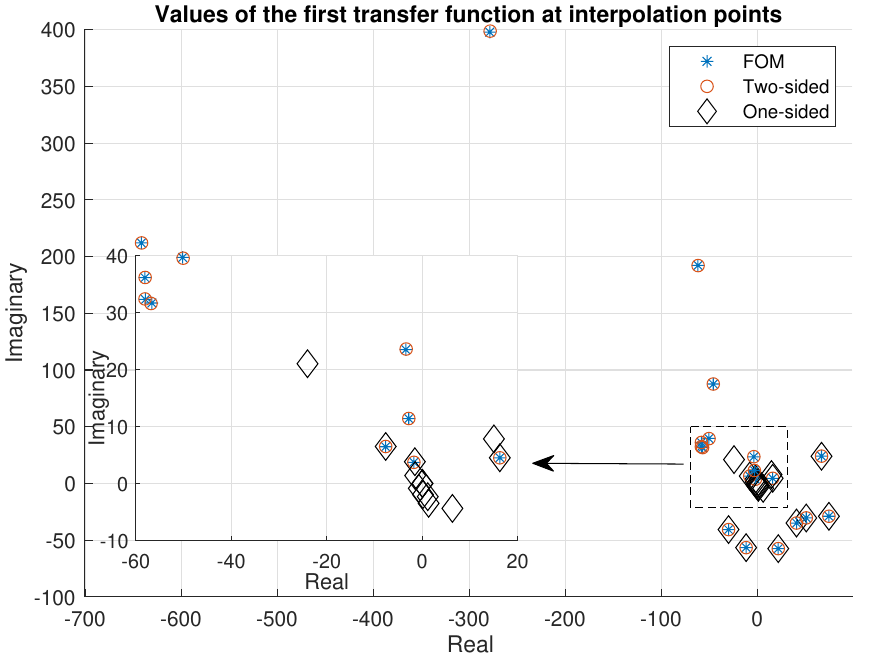}
      \caption{Interpolation conditions for the first component of the transfer matrix. The blue stars, red circles and black diamonds represent the quantities related to the FOM, ROM (two-sided) and ROM (one-sided), respectively.}
      \label{MomentValue1}
\end{figure}

\begin{figure}[htpb]
      \centering
      \includegraphics[width=0.9\columnwidth]{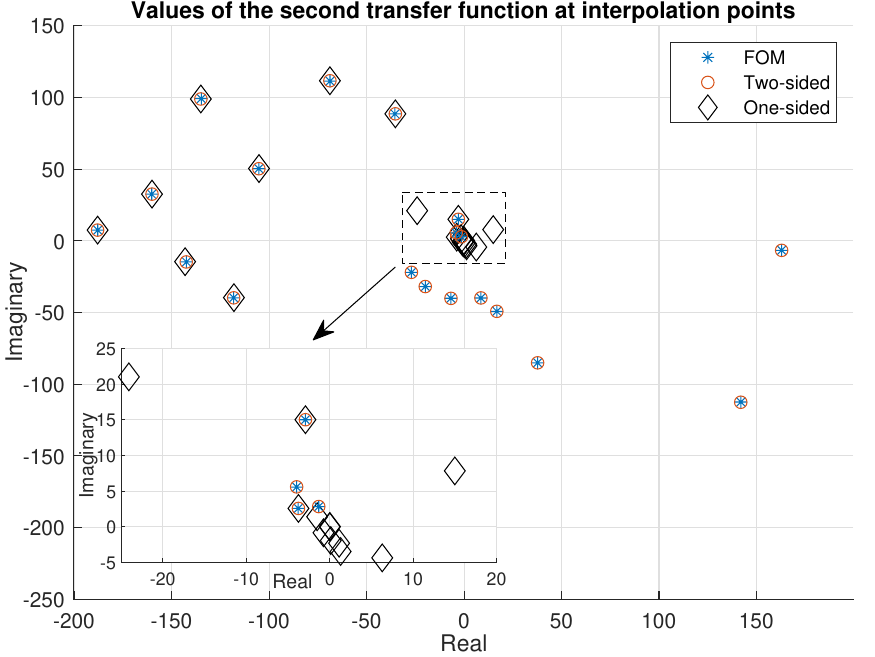}
      \caption{Interpolation conditions for the second component of the transfer matrix. The blue stars, red circles and black diamonds represent the quantities related to the FOM, ROM (two-sided) and ROM (one-sided), respectively.}
      \label{MomentValue2}
\end{figure}

Furthermore, to illustrate the two-sided property, the proposed method is compared with the one-sided method given by \cite{scarciotti2016low}. Since the FOM in this example has two inputs and two outputs, the moments along the tangential directions consist of two transfer functions (see Definition~\ref{D1} and Definition~\ref{D2}). The values of the first and second transfer function at the required interpolation points are presented with blue stars in Fig.~\ref{MomentValue1} and Fig.~\ref{MomentValue2}, respectively. These figures confirm that for the constructed ROMs (represented by red circles for the two-sided method and black diamonds for the method in\cite{scarciotti2016low}) of the same order $20$, the proposed two-sided method ensures double interpolations. In fact, the two-sided method achieves moment matching at all $40$ interpolation points (all blue stars are surrounded by red circles), while the one-sided method achieves the moment matching at $20$ interpolation points only ($20$ black diamonds surround $20$ blue stars, but other $20$ diamonds are empty, meaning that $20$ prescribed interpolation points are missed).


In conclusion, in terms of quality of matching, the performance of the data-driven algorithm in absence of noise is satisfactory and equivalent to a model-based method. It is important to stress that this result is very valuable even when the model of the system is available and noiseless data is generated by means of simulations because, as shown later in Section~\ref{Sec_time}, the data-driven method produces a reduced-order model in much less computational time than the model-based approach. Nevertheless, we study the robustness of the data-driven method to measurement noise in the next section.

\subsection{Robustness against measurement noise}\label{Sec_Robust}
\begin{figure}[thpb]
      \centering
      \includegraphics[width=0.9\columnwidth]{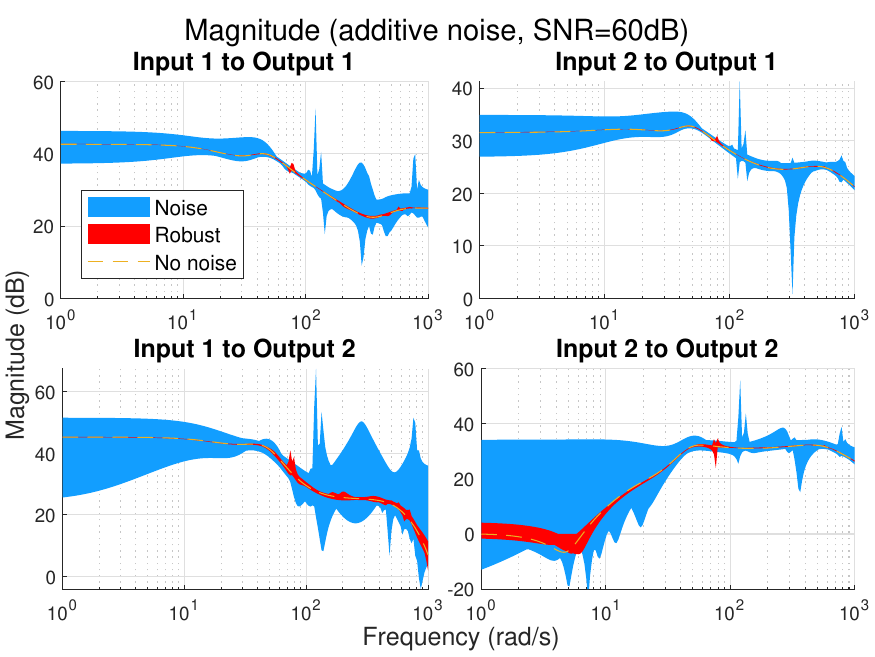}
      \caption{Magnitude plots of the reduced-order models under 20 realizations of additive Gaussian noise. The red regions, blue regions and brown dashed lines refer to the robust algorithm (Algorithm~\ref{algYB}), non-robust method (Theorem~\ref{swapped_dd}) and the ideal estimation (no noise), respectively.}
      \label{sim_noise}
\end{figure}

To evaluate the robustness of the proposed data-driven algorithm, we add Gaussian noise to the measurements of $\varpi(t)$. We consider $20$ realisations of the noise with signal-noise-ratio (SNR) of $60$dB, and we estimate the quantity $\Upsilon B$ both by the robust (Algorithm~\ref{algYB}) and the non-robust (Theorem~\ref{swapped_dd}) method\footnote{We have carried out the same analysis adding noise to $y(t)$ and studying the robustness of Algorithm~\ref{algCPai} to estimate $C\Pi$ (using sufficiently large $\widetilde{\nu}$). {The performance of Algorithm~\ref{algCPai} and Algorithm~\ref{algYB} under a larger level of noise (\emph{i.e.} $40$dB) is also evaluated. These} results are analogous and thus they are omitted for the reason of space. However, these can be found in\cite{appendix}.}. The average of $e_{\Upsilon B}$ over $20$ realisations is denoted as $\bar e_{\Upsilon B}^r$ for the robust and $\bar e_{\Upsilon B}$ for the non-robust estimation. The corresponding average of $e_{\Upsilon\Pi}$ is denoted as $\bar e_{\Upsilon\Pi}^r$ and $\bar e_{\Upsilon\Pi}$, respectively. The robust algorithm (Algorithm~\ref{algYB}) uses $1000$ measurements, while the non-robust method (Theorem~\ref{swapped_dd}) uses only one measurement (the last). The robust estimations have small averaged errors, \textit{i.e.} $\bar e_{\Upsilon B}^r=3.2\times 10^{-5}$ and $\bar e_{\Upsilon\Pi}^r=5.3\times 10^{-5}$, while the non-robust estimations present averaged errors two orders of magnitude higher, namely $\bar e_{\Upsilon B}=1.0\times 10^{-3}$ and $\bar e_{\Upsilon\Pi}=1.4\times 10^{-3}$, respectively. Fig.~\ref{sim_noise} shows the regions between the upper and lower envelopes of all obtained magnitude plots of the ROMs obtained for the $20$ realisations of the noise. In particular, the red regions refer to the robust estimation (Algorithm~\ref{algYB}), the blue regions refer to the non-robust estimation (Theorem~\ref{swapped_dd}) and the brown dashed lines represent the ideal estimation (no noise). This figure confirms that the proposed data-driven algorithm (Algorithm~\ref{algYB}) is robust with respect to measurement noise as, for $\widetilde{q}$ sufficiently large, the red regions are not only significantly smaller than the blue regions, but they are also close to the ideal case.  

\subsection{Initial validation: stability and PCC voltage}\label{Sec_validation}
\begin{figure}[thpb]
      \centering
      \includegraphics[width=0.9\columnwidth]{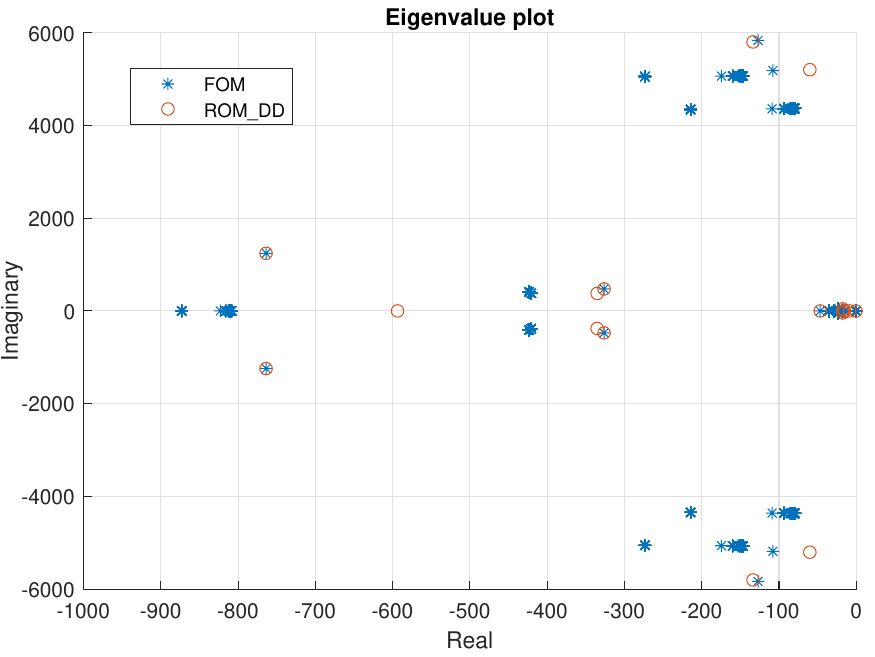}
      \caption{Eigenvalues of the FOM and ROM\_DD (data-driven), represented by the blue stars and hollow red circles, respectively.}
      \label{eigplot}
\end{figure}

\begin{figure}[thpb]
      \centering
      \includegraphics[width=0.9\columnwidth]{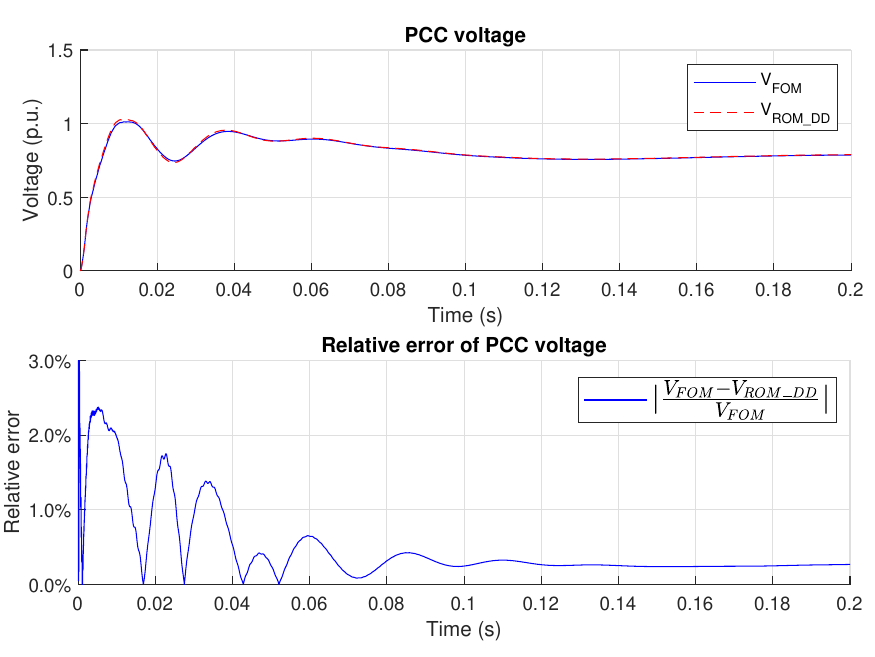}
      \caption{Top: PCC voltage of the FOM and ROM\_DD (data-driven), represented by solid blue and dashed red lines, respectively. Bottom: corresponding relative error.}
      \label{Pcc}
\end{figure}

In the two-sided approach, the free parameter $G$ is used to double the number of interpolation points (See Section~\ref{Sec_twosided} and \eqref{rom_twos}). We select the interpolation points (encoded by $S$) and tangential directions $L$ such that the ROM is stable. For further validation, we first compare the eigenvalues of the ROM and FOM and then we reconnect them with the study area to compare the PCC voltage. In Fig.~\ref{eigplot}, the eigenvalues of the FOM are represented by blue stars while the eigenvalues of the ROM obtained by the data-driven MOR method are illustrated by hollow red circles. The figure shows that the obtained ROM is stable and that the eigenvalues of the ROM have a similar spread as those of the FOM. Fig.~\ref{Pcc} compares the PCC voltage of the FOM and ROM when they are reconnected to the study area. The top plot shows the PCC voltages in per-unit (with $V_{base}=66\rm kV$), whereas the bottom plot shows the relative error in percentage. The relative error of the PCC voltage decreases to less than $0.3\%$ within $0.1s$. Hence, the simulations confirm that the ROM can accurately recover the performance of the FOM in a short time. 

\subsection{Comparison of computational time}\label{Sec_time}
In order to evaluate the advantages of the data-driven MOR approach proposed in this paper, we compare the computational time of four wind farm models: 1) the nonlinear FOM (of the external area), 2) its linearisation, 3) the ROM constructed by solving the Sylvester equation (\textit{i.e.} model-based method) and 4) the ROM constructed by the proposed data-driven approach. In addition, to show that the proposed data-driven method is advantageous in the synthetic data scenario, we compare also the ``extraction time'' of the MOR (that is, the time required to construct the ROM) for the data-driven method and the model-based method. Recall that the FOM of the wind farm has order $3402$. The two ROMs have order $20$. 

Table~\ref{comparison} compares the computational time of the four models in terms of ``extraction time'' of the MOR and ``elapsed time'' of a $1$ second simulation of the wind farm. All the simulations and MOR algorithms are run in {Matlab/Simulink} R2023b on the same personal laptop (Windows 10, Intel Core i7-10750H and 16GB RAM). The table shows that to perform a simulation of $1$ second, the nonlinear FOM requires $1268.26$ seconds (about $21$ minutes), whereas the linearised FOM requires $796.31$ seconds (about $13$ minutes), which is time-consuming. By reducing the order of the model from $3402$ to $20$, the elapsed time is significantly reduced to around $1.5$ seconds for both the model-based ROM and the data-driven ROM. While the elapsed time of the two ROMs is similar, the two methods show a stark difference in the extraction time, which reflects the computational complexity of the MOR methods. In fact, to construct a ROM of order $20$ using a model-based approach requires $28.60$ seconds, whereas using the data-driven method takes only $0.10$ seconds. In summary, a simulation using the data-driven ROM is $530$ times faster than a simulation using the linearised FOM, and this model is $280$ times faster to construct than a model-based ROM. Note that the difference in extraction time can be justified theoretically and is independent of the specific model-based method used. In fact, any model-based ROM method operates on the matrices $A$, $B$, $C$ which have order $n$, whereas the proposed algorithms have computational complexity that is independent of $n$ and depends only on $\nu$. As $\nu \ll n$, the proposed method is faster. 

\begin{table}[htbp]
\caption{Computational comparison}
\centering
\begin{tabular}{|c|c|c|c|}
\hline
Model&Order&Extraction time&Elapsed time\\
\hline
FOM (nonlinear)&3402&-&1268.26s\\
\hline
FOM (linear)&3402&-&796.31s\\
\hline
ROM&20&28.60s&1.53s\\
\hline
ROM\_DD&20&0.10s&1.52s\\
\hline
\end{tabular}
\label{comparison}
\end{table}

\subsection{Advanced validation: fault behaviour}\label{Sec_fault}

Further validation is provided by means of fault simulations. Two fault scenarios are considered:
\begin{itemize}
    \item {1) A self-clearing fault at bus 14 of the study area occurring at $t=0.3\rm s$ and cleared after $\rm 50ms$.}
    \item {2) A fault at bus 8 of the study area occurring at $t=0.3\rm s$. After $\rm 100ms$ bus 8 is disconnected to protect the rest of the power system.}
\end{itemize}
All the faults are produced by the `Three-Phase Fault' block in the Simulink.

In this section, we reconnect the external area to the study area to verify if the ROM of the external area behaves well not just in isolation, but in the realistic case in which this is interconnected to the rest of the power system. For both cases, the PCC voltage is presented in per-unit (with $V_{base}=66\rm kV$) while the corresponding relative error is shown in percentage.

\subsubsection{Case study 1 (Self-clearing fault)}

\begin{figure}[thpb]
      \centering
      \includegraphics[width=0.9\columnwidth]{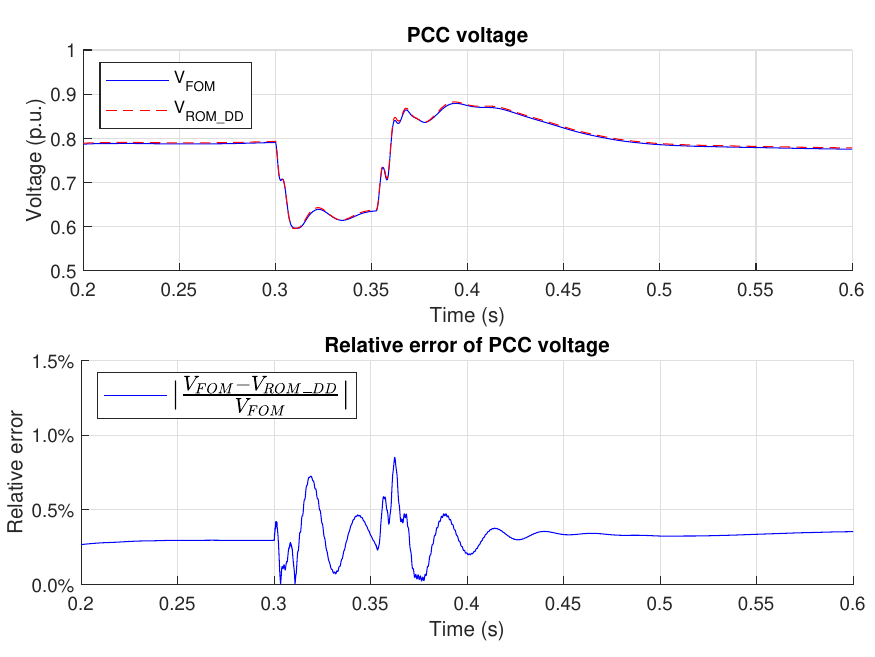}
      \caption{Top: PCC voltage of the FOM and ROM\_DD (data-driven), represented by solid blue and dashed red lines, respectively. Bottom: corresponding relative error.}
      \label{fault1}
\end{figure}

A fault occurs at bus 14 of the power system at $t=0.3\rm s$ and is cleared at $t=0.35\rm s$. The PCC voltage drops by $19.43\%$ and recovers within $\rm 0.15s$ after the fault is cleared. The top of Fig.~\ref{fault1} shows the PCC voltage of the study area when it is reconnected to the FOM (solid blue lines) and the ROM (dashed red lines) respectively. The PCC voltage of the two models is almost indistinguishable when the bus 14 experiences the fault. The bottom of Fig.~\ref{fault1} shows the respective relative errors (in percentage). The largest relative error occurs in the transient period after the PCC voltage drops or steps (\emph{i.e.} $\rm t=0.32s$ and $t=0.36s$). These peak errors are small ($0.72\%$ and $0.84\%$) and decrease to less than $0.4\%$ within $0.1s$. We conclude that when a self-clearing fault occurs at the bus of the power system, the influence on the PCC voltage is recovered precisely by the ROM. Thus, the fault behaviour is fully preserved by the data-driven MOR approach.

\subsubsection{Case study 2 (Not self-clearing fault)}

\begin{figure}[thpb]
      \centering
      \includegraphics[width=0.9\columnwidth]{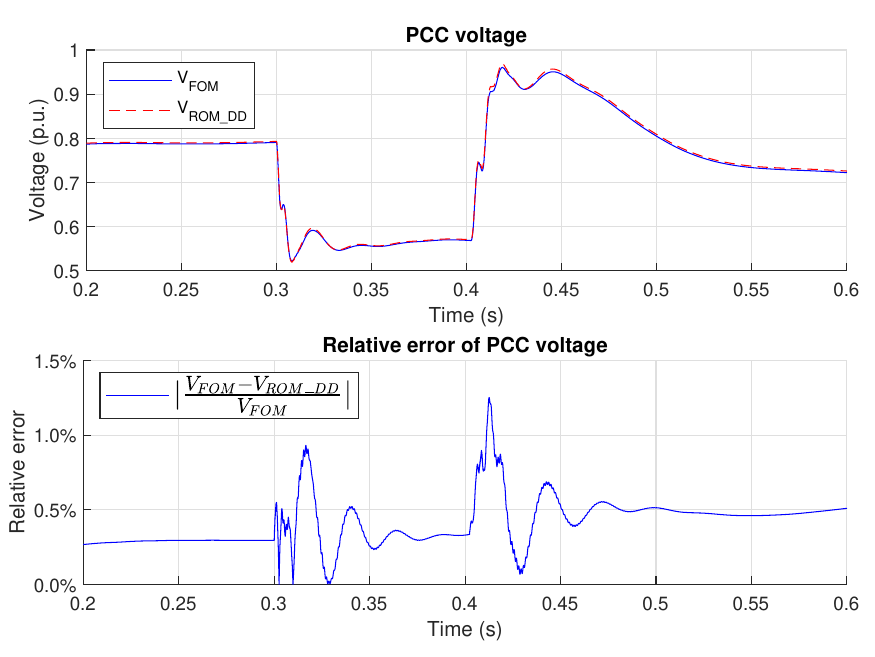}
      \caption{Top: PCC voltage of the FOM and ROM\_DD (data-driven), represented by solid blue and dashed red lines, respectively. Bottom: corresponding relative error.}
      \label{fault2}
\end{figure}

A fault occurs at bus 8 of the power system at $t=0.3\rm s$ and cannot be cleared. Bus 8 is disconnected at $t=0.4\rm s$ to protect the rest of the power system. This fault results in a voltage drop of $28.04\%$ at the PCC. The top of Fig.~\ref{fault2} shows the PCC voltage of the combined system of the study area and the FOM (solid blue lines) or ROM (dashed red lines), respectively. The bottom of Fig.~\ref{fault2} shows the corresponding relative errors of the PCC voltage with a peak of $1.22\%$ at $t=0.41\rm s$ and decreasing to less than $0.5\%$ within $0.1s$. Since this fault cannot be self-cleared, it is isolated by disconnecting bus 8. Missing the generator at bus 8, the PCC voltage cannot recover and decreases to a smaller value of $\rm 0.73pu$. This behavior is fully captured by the use of the ROM in place of the FOM.  

\begin{figure}[thpb]
      \centering
      \includegraphics[width=0.9\columnwidth]{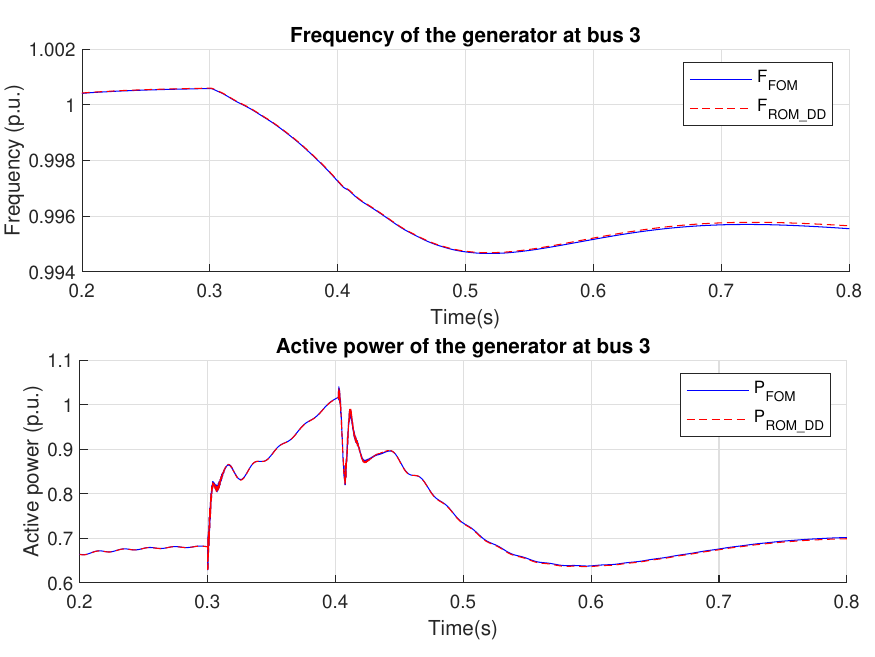}
      \caption{Top: Frequency of the generator at bus 3. Bottom: Active power of the generator at bus 3. The solid blue and dashed red lines represent the quantities related to the FOM and ROM\_DD (data-driven), respectively.}
      \label{Generator3}
\end{figure}

Moreover, we further evaluate the influence of the ROM on the study area in terms of generator frequency and active power. We consider the same fault scenario as described in the case study 2. The top of Fig.~\ref{Generator3} shows the generator frequency at bus 3 of the study area (with $f_{base}=50 \rm Hz$). This frequency drops down at $t=0.3\rm s$ and finally returns to around $\rm0.996pu$ after the fault is isolated. The bottom of Fig.~\ref{Generator3} shows the active power at bus 3 of the study area (with $P_{base}=100 \rm MW$). When the fault occurs, the active power steps to more than $\rm1.0pu$ within $0.1\rm s$  and finally returns to $\rm0.70pu$ at the end. Fig.~\ref{Generator3} confirms that the performance of the FOM when reconnected to the study area is fully recovered by the ROM.

Combining all the simulation results presented in this section, we conclude that by reducing the external area from a complicated FOM to a simple ROM, we considerably reduce the elapsed time of simulations while maintaining high fidelity across a range of different criteria. In addition, the proposed method is robust to measurement noise, it is completely model free and has the advantage of low computational complexity (\textit{i.e.} small extraction time) even in the synthetic data scenario.

\section{Conclusion}\label{Sec_conclusion}

We have presented a data-driven algorithm for the MOR of large-scale wind farms. We presented algorithms to achieve two-sided moment matching for MIMO systems. Unlike other MOR approaches that require full information of the system matrices, the proposed data-driven algorithm is model free, robust to measurement noise and offers low computational complexity to practically overcome the challenges posed by large-scale wind farm. Finally, we have provided extensive simulations that validate the proposed method on a
combined model of a 200-turbine wind farm interconnected to the IEEE
14-bus system in terms of Bode plot, eigenvalues, PCC voltage and fault behaviour.

\bibliographystyle{IEEEtran}
\bibliography{bibliography}

\begin{thebibliography}{10}
\providecommand{\url}[1]{#1}
\csname url@samestyle\endcsname
\providecommand{\newblock}{\relax}
\providecommand{\bibinfo}[2]{#2}
\providecommand{\BIBentrySTDinterwordspacing}{\spaceskip=0pt\relax}
\providecommand{\BIBentryALTinterwordstretchfactor}{4}
\providecommand{\BIBentryALTinterwordspacing}{\spaceskip=\fontdimen2\font plus
\BIBentryALTinterwordstretchfactor\fontdimen3\font minus \fontdimen4\font\relax}
\providecommand{\BIBforeignlanguage}[2]{{%
\expandafter\ifx\csname l@#1\endcsname\relax
\typeout{** WARNING: IEEEtran.bst: No hyphenation pattern has been}%
\typeout{** loaded for the language `#1'. Using the pattern for}%
\typeout{** the default language instead.}%
\else
\language=\csname l@#1\endcsname
\fi
#2}}
\providecommand{\BIBdecl}{\relax}
\BIBdecl

\bibitem{diaz2020review}
H.~D{\'\i}az and C.~G. Soares, ``Review of the current status, technology and future trends of offshore wind farms,'' \emph{Ocean Eng.}, vol. 209, p. 107381, 2020.

\bibitem{usdata}
{U.S. Energy Information Administration}, ``Monthly energy review,'' https://www.eia.gov/todayinenergy/detail.php?id=43675, 2020.

\bibitem{scarciotti2016low}
G.~Scarciotti, ``Low computational complexity model reduction of power systems with preservation of physical characteristics,'' \emph{IEEE Trans. Power Syst.}, vol.~32, no.~1, pp. 743--752, 2016.

\bibitem{ali2018model}
H.~R. Ali, L.~P. Kunjumuhammed, B.~C. Pal, A.~G. Adamczyk, and K.~Vershinin, ``Model order reduction of wind farms: Linear approach,'' \emph{IEEE Trans. Sustain. Energy}, vol.~10, no.~3, pp. 1194--1205, 2018.

\bibitem{ali2019trajectory}
------, ``A trajectory piecewise-linear approach to nonlinear model order reduction of wind farms,'' \emph{IEEE Trans. Sustain. Energy}, vol.~11, no.~2, pp. 894--905, 2019.

\bibitem{ali2020model}
H.~R. Ali and B.~C. Pal, ``Model order reduction of multi-terminal direct-current grid systems,'' \emph{IEEE Trans. Power Syst.}, vol.~36, no.~1, pp. 699--711, 2020.

\bibitem{akhmatov2002aggregate}
V.~Akhmatov and H.~Knudsen, ``An aggregate model of a grid-connected, large-scale, offshore wind farm for power stability investigations—importance of windmill mechanical system,'' \emph{Int. J. Electr. Power Energy Syst.}, vol.~24, no.~9, pp. 709--717, 2002.

\bibitem{slootweg2003aggregated}
J.~Slootweg and W.~Kling, ``Aggregated modelling of wind parks in power system dynamics simulations,'' in \emph{2003 IEEE Bologna Power Tech. Conf. Proc.}, vol.~3.\hskip 1em plus 0.5em minus 0.4em\relax IEEE, 2003, pp. 6--pp.

\bibitem{fernandez2008aggregated}
L.~M. Fern{\'a}ndez, F.~Jurado, and J.~R. Saenz, ``Aggregated dynamic model for wind farms with doubly fed induction generator wind turbines,'' \emph{Renew. Energy}, vol.~33, no.~1, pp. 129--140, 2008.

\bibitem{fernandez2009equivalent}
L.~Fernandez, C.~Garcia, J.~Saenz, and F.~Jurado, ``Equivalent models of wind farms by using aggregated wind turbines and equivalent winds,'' \emph{Energy Convers. Manage.}, vol.~50, no.~3, pp. 691--704, 2009.

\bibitem{ali2012wind}
M.~Ali, I.-S. Ilie, J.~V. Milanovic, and G.~Chicco, ``Wind farm model aggregation using probabilistic clustering,'' \emph{IEEE Trans. Power Syst.}, vol.~28, no.~1, pp. 309--316, 2012.

\bibitem{zou2015fuzzy}
J.~Zou, C.~Peng, H.~Xu, and Y.~Yan, ``A fuzzy clustering algorithm-based dynamic equivalent modeling method for wind farm with dfig,'' \emph{IEEE Trans. Energy Convers.}, vol.~30, no.~4, pp. 1329--1337, 2015.

\bibitem{wang2018improved}
P.~Wang, Z.~Zhang, Q.~Huang, N.~Wang, X.~Zhang, and W.-J. Lee, ``Improved wind farm aggregated modeling method for large-scale power system stability studies,'' \emph{IEEE Trans. Power Syst.}, vol.~33, no.~6, pp. 6332--6342, 2018.

\bibitem{moore1981principal}
B.~Moore, ``Principal component analysis in linear systems: Controllability, observability, and model reduction,'' \emph{IEEE Trans. Autom. Control}, vol.~26, no.~1, pp. 17--32, 1981.

\bibitem{gugercin2004survey}
S.~Gugercin and A.~C. Antoulas, ``A survey of model reduction by balanced truncation and some new results,'' \emph{Int. J. Control}, vol.~77, no.~8, pp. 748--766, 2004.

\bibitem{antoulas2005approximation}
A.~C. Antoulas, \emph{Approximation of large-scale dynamical systems}.\hskip 1em plus 0.5em minus 0.4em\relax SIAM, 2005.

\bibitem{grimme1997krylov}
E.~J. Grimme, ``Krylov projection methods for model reduction,'' Ph.D. dissertation, University of Illinois at Urbana-Champaign, 1997.

\bibitem{druskin2011adaptive}
V.~Druskin and V.~Simoncini, ``Adaptive rational krylov subspaces for large-scale dynamical systems,'' \emph{Syst. \& Control Letter.}, vol.~60, no.~8, pp. 546--560, 2011.

\bibitem{chaniotis2005model}
D.~Chaniotis and M.~Pai, ``Model reduction in power systems using krylov subspace methods,'' \emph{IEEE Trans. Power Syst.}, vol.~20, no.~2, pp. 888--894, 2005.

\bibitem{freitas2008gramian}
F.~D. Freitas, J.~Rommes, and N.~Martins, ``Gramian-based reduction method applied to large sparse power system descriptor models,'' \emph{IEEE Trans. Power Syst.}, vol.~23, no.~3, pp. 1258--1270, 2008.

\bibitem{chow2013power}
J.~H. Chow, \emph{Power system coherency and model reduction}.\hskip 1em plus 0.5em minus 0.4em\relax Springer, 2013, vol.~84.

\bibitem{sturk2014coherency}
C.~Sturk, L.~Vanfretti, Y.~Chompoobutrgool, and H.~Sandberg, ``Coherency-independent structured model reduction of power systems,'' \emph{IEEE Trans. Power Syst.}, vol.~29, no.~5, pp. 2418--2426, 2014.

\bibitem{druskin2011analysis}
V.~Druskin, L.~Knizhnerman, and V.~Simoncini, ``Analysis of the rational krylov subspace and adi methods for solving the lyapunov equation,'' \emph{SIAM J. Numeri. Anal.}, vol.~49, no.~5, pp. 1875--1898, 2011.

\bibitem{gugercin2006rational}
S.~Gugercin, C.~Beattie, and A.~Antoulas, ``Rational krylov methods for optimal $h_2$ model reduction,'' in \emph{17th Int. Symp. Math. Theory Networks Syst. (MTNS)}, vol. 2006, no.~7, 2006, pp. 1665--1667.

\bibitem{scarciotti2017data}
G.~Scarciotti and A.~Astolfi, ``Data-driven model reduction by moment matching for linear and nonlinear systems,'' \emph{Automatica}, vol.~79, pp. 340--351, 2017.

\bibitem{shakib2023time}
M.~F. Shakib, G.~Scarciotti, A.~Y. Pogromsky, A.~Pavlov, and N.~van~de Wouw, ``Time-domain moment matching for multiple-input multiple-output linear time-invariant models,'' \emph{Automatica}, vol. 152, p. 110935, 2023.

\bibitem{mao2022data}
J.~Mao and G.~Scarciotti, ``Data-driven model reduction by moment matching for linear systems through a swapped interconnection,'' in \emph{2022 Euro. Control Conf. (ECC)}.\hskip 1em plus 0.5em minus 0.4em\relax IEEE, 2022, pp. 1690--1695.

\bibitem{ionescu2015two}
T.~C. Ionescu, ``Two-sided time-domain moment matching for linear systems,'' \emph{IEEE Trans. Autom. Control}, vol.~61, no.~9, pp. 2632--2637, 2015.

\bibitem{mao2024data}
J.~Mao and G.~Scarciotti, ``Data-driven model reduction by two-sided moment matching,'' \emph{Automatica}, vol. 166, p. 111702, 2024.

\bibitem{appendix}
Z.~Gong, J.~Mao, A.~Junyent-Ferr\'{e}, and G.~Scarciotti, ``Appendix: Detailed wind farm model and {IEEE} 14-bus system,'' {https://github.com/zilong-gong/WindFarmMOR}, 2024.

\bibitem{junyent2011control}
A.~Junyent~Ferr{\'e}, ``Control of power electronic converters for the operation of wind generation,'' Ph.D. dissertation, Universitat Polit{\`e}cnica de Catalunya, 2011.

\bibitem{padoan2017geometric}
A.~Padoan, G.~Scarciotti, and A.~Astolfi, ``A geometric characterization of the persistence of excitation condition for the solutions of autonomous systems,'' \emph{IEEE Trans. Autom. Control}, vol.~62, no.~11, pp. 5666--5677, 2017.

\end{thebibliography}

\appendix

\section*{A. Proofs of the technical results}\label{Appendix proof}

In this section, we provide the proofs of the theorems presented in the paper.

\subsection*{Proof of Theorem~\ref{direct_dd}}
The minimality of the triple $(L,S,\omega(0))$ implies the excitability of the pair $(S,\omega(0))$. By \cite[Lemma 3]{padoan2017geometric}, there exists a time sequence $\{t_k\}$ with $\lim_{k\to \infty}t_k=\infty$ such that the time-snapshots $\widetilde{R}_k$ is column full rank. The rest of the proof is the same as the proof of \cite[Theorem 2]{scarciotti2016low}. \ \hfill $\square$ 

\subsection*{Proof of Theorem~\ref{swapped_dd}}
Since $\Upsilon$ is the unique solution of the Sylvester equation \eqref{syl2}, we have
\begin{equation}
\label{d}
\begin{aligned}
\dot{d}&=\dot{\varpi}+\Upsilon\dot{x}=Q\varpi+Ry+\Upsilon(Ax+Bu)\\
&=Q\varpi+(RC+\Upsilon A)x+\Upsilon Bu=Q(\varpi+\Upsilon x)+\Upsilon Bu\\
&=Qd+\Upsilon Bu
\end{aligned}
\end{equation}
Since the initial condition $x(0)=0$ and $\varpi(0)=0$, we have $d(0)=\varpi(0)+\Upsilon x(0)=0$. Then, the input $u=e_j\delta_0(t)$ yields $\varpi(t)+\Upsilon x(t)=d(t)=e^{Qt}\Upsilon B^{j}$ for all $t>0$. Note that system \eqref{system} is assumed to be asymptotically stable. Thus, $x(t)\to 0$ as $t\to \infty$ yielding
$$
\begin{aligned}
\Upsilon B^{j}&=\lim_{t\to \infty}e^{-Qt}(\varpi(t)+\Upsilon x(t))\\
&=\lim_{t\to \infty}e^{-Qt}\varpi(t).\\
\end{aligned}
$$
This proves Theorem~\ref{swapped_dd}.
\ \hfill $\square$

The robustness claim is a consequence of \eqref{YBls} being the least-square version of \eqref{YB}.

\section*{B. model and parameters}\label{Appendix model}
In this section, we summarise the main parameters used in the simulation. The detailed model and parameters are available in\cite{appendix}.

\section*{Nomenclature}

\textbf{Symbols}

\begin{tabular}{l@{\quad}l}
$\textit{A}$ & Wind turbine swept surface\\
$C$ & DC bus capacity\\
$I_t$ & One-mass wind turbine aggregated inertia\\
$L_l$ & Inductance of the grid connection impedance\\
$L_q$ & Generator inductance on $q$ axis\\
$L_d$ & Generator inductance on $d$ axis\\
$R$ & Wind turbine radius\\
$V_{DC}^*$ & DC bus voltage reference value\\
$f^r$ & Rated frequency\\
$r_s$ & Resistance of a single phase of the stator windings\\
$r_l$ & Resistance of the grid connection impedance\\
$U^r_g$ & Rated collection grid voltage\\
$U^r_w$ & Rated wind turbine output voltage\\
$v_w$ & Wind speed\\
$\lambda_m$ & Flux linkage per rotating speed unit\\
$\nu$ & Gearbox multiplication ratio\\
$\omega_{mn}$ & Nominal generator speed\\
$\rho$ & Air density\\
$\tau$ & Time constant of the pitch angle controller
\end{tabular}

\hspace*{\fill}

\textbf{Controller gains}

\begin{tabular}{l@{\qquad}l}
$K_p$, $K_i$ & gains of the turbine speed controller\\
$K_{pq}$, $K_{iq}$ & gains of the generator controller for $q$ axis\\
$K_{pd}$, $K_{id}$ & gains of the generator controller for $d$ axis\\
$K_{pg}$, $K_{ig}$ & gains of the grid side controller\\
$K_{pc}$, $K_{ic}$ & gains of the grid current controller
\end{tabular}

\section*{Wind Farm Parameters}

\noindent1) Wind turbine (see \cite[Chapter 2]{junyent2011control} for details): $c_1=1$, $c_2=39.52$, $c_6=2.04$, $c_7=14.47$, $c_3=c_4=c_5=c_8=c_9=0$, $R=40$m, $\emph{A}=5,026.5\rm m^2$, $\rho=1.225\rm kg/ m^3$, $\nu=90$, $I_t=4\rm kg\cdot km^2$, $\tau=0.1\rm s$ and $v_w=7\rm m/s$.

\noindent2) Wind turbine speed controller: $\omega_{mn}=1,602\rm min^{-1}$, $K_p=0.1^{\circ}\cdot s\rm/rad$ and $K_i=0.02^{\circ}\rm/rad$.

\noindent3) Generator: 2 pairs of poles, $r_s=15\rm m\Omega$, $\lambda_m=2.35\rm V\cdot s/rad$, $L_q=0.12732$mH and $L_d=0.12764$mH.

\noindent4) Generator vector controller: $K_{pq}=0.0637\rm V/A$, $K_{iq}=7.5\rm V/(A\cdot s)$, $K_{pd}=0.0638\rm V/A$ and $K_{id}=7.5\rm V/(A\cdot s)$.

\noindent5) DC bus: $C=10$mF and $V_{DC}^*=2.6\rm kV$.

\noindent6) Grid side system: $r_l=20\rm m\Omega$, $L_l=1\rm mH$, $U^{r}_w=0.97\rm kV$, $U^{r}_g=66\rm kV$ and $f^{r}=50\rm Hz$.

\noindent7) Grid side controller: $K_{pg}=0.6032\rm A/V$ and $K_{ig}=14.2122\rm A/(V\cdot s)$.

\noindent8) Grid current controller: $K_{pc}=0.2803\rm V/A$ and $K_{ic}=10\rm V/(A\cdot s)$. 

\section*{C. Code availability}\label{Appendix Demo}
A demo of the method is available at \url{https://github.com/zilong-gong/WindFarmMOR}.




\end{document}